\documentclass[aps,twocolumn,groupedaddress]{revtex4-1}
\usepackage{graphicx}

\begin{document}

%\preprint{}

\title{Observation of Multiple Superconducting Gaps\\ in the Infrared Reflectivity Spectra of Ba(Fe$_{0.9}$Co$_{0.1}$)$_2$As$_2$}



%\author{Yu. A. Aleshchenko}
%\email{yuriale@sci.lebedev.ru}
%\thanks{}
%\altaffiliation{ }
%\affiliation{}

\author{Yu. A. Aleshchenko}
	\email{yuriale@sci.lebedev.ru}
\author{A. V. Muratov}
%	\email{muratov@sci.lebedev.ru}
\author{V. M. Pudalov}
\affiliation{Lebedev Physical Institute, Russian Academy of Sciences, Moscow, 119991 Russia}

\author{E. S. Zhukova}
\author{B. P. Gorshunov}
\affiliation{Prokhorov General Physics Institute, Russian Academy of Sciences, Moscow, 119991 Russia}
\affiliation{Moscow Institute of Physics and Technology, Dolgoprudnyi, Moscow region, 141700 Russia}

\author{F. Kurth}
\author{K. lida}
\affiliation{Institut f\"ur Metallische Werkstoffe, Leibniz-Instilut f\"ur Fesik\"orper- und Werkstoffforschung Dresden,
D-01171 Dresden, Germany }

\date[Date: ]{20 October 2011}

\begin{abstract}
The results of infrared reflectivity measurements for the iron-based high-temperature superconductor $\mathrm{Ba(Fe_{0.9}Co_{0.1})_2As_2}$ are reported. The reflectivity is found to be close to unity at frequencies $\omega$ lower than $2\Delta/h$ ($2\Delta$ is the superconducting gap and $h$ is Planck's constant). This is evidence for the $s^{+/-}$ or $s^{+/+}$ symmetry of the superconducting order parameter in the studied compound. The infrared reflectivity spectra of $\mathrm{Ba(Fe_{0.9}Co_{0.1})_2As_2}$ manifest opening of several superconducting gaps at temperatures lower than critical $T_c$.
\end{abstract}

%\keywords{}

\maketitle

Studies of the recently discovered iron-based high-temperature superconductors (HTSC) \cite{1} and comparison of their properties with those for cuprate HTSC provide information necessary to understand the mechanisms of high-temperature superconductivity. The energy gap, 2$\Delta$, in the spectrum of quasi-particle excitations is one of the most important parameters of a superconductor. For iron-based compounds, experimental studies of this parameter, as well as the mechanisms of superconductivity, are hampered by the multiband structure of their energy spectrum. The Brillouin zone in these materials contains two hole bands at the $\Gamma$ point and two electron bands at the X(0, $\pm\pi$) points. Therefore, we can expect the existence in this material of several superconducting (SC) condensates with different energy gaps \cite{2,3,4}.

The complexity of the energy spectrum in iron pnictides, the difference in quality and phase composition of the earlier studied samples, significant experimental errors, as well as the different sensitivity of various experimental techniques to the different energy gaps may be responsible for wide scattering in the measured ratios of the SC energy gap to the critical temperature, $2\Delta/k_BT_c = 1.6-10$ ($k_B$ is the Boltzmann constant) \cite{5,6,7,8,9,10,11,12}. The reliable data on the symmetry of the order parameter in these materials are also not available yet. Within the framework of the most widely discussed s$^{+/-}$ model of superconductivity \cite{13} it is assumed that the SC condensates of the electron and hole bands in iron-based HTSC possess the $s-$type order parameters with the Bardeen-Cooper-Schrieffer (BCS)-like temperature dependence and with the opposite sign. However, different conclusions have been reported in various experiments including observations of the SC energy gap with $s-$type symmetry \cite{14,15} and with nodes at the Fermi surface, where the order parameter changes sign, i.e., with the $d-$type symmetry \cite{8,16,17,18}.

The infrared (IR) spectroscopy is a direct technique to gain information on the energy spectrum of charge carriers in superconductors. Contrary to angle-resolved photoemission spectroscopy (ARPES) and tunnel spectroscopy, a relatively thick layer of material contributes to the IR reflectivity, thus making the observed properties similar to those of the bulk material. However, studies of iron pnictides by the latter technique face with obstacles because typical magnitude of the superconducting gaps (at $T_c \sim 20-55$~K) lies in the far-IR and terahertz (THz) ranges. The efficiency of standard IR Fourier spectrometers in these spectral ranges decreases dramatically. In this paper, we report studies of the IR reflectivity spectra of Ba(Fe$_{0.9}$Co$_{0.1}$)$_2$As$_2$ film in the wide ranges of wavelengths and temperatures. The high optical efficiency of the Bruker IFS 125HR Fourier spectrometer and the high sensitivity of its detectors enabled us to overcome the above obstacles and to perform measurements even in the far-IR range.

The Ba(Fe$_{0.9}$Co$_{0.1}$)$_2$As$_2$ film was deposited on a $4 \times 9$ mm$^2$ (La, Sr)(Al, Ta)O$_3$ substrate by pulsed laser deposition technique, where the Ba(Fe$_{0.9}$Co$_{0.1}$)$_2$As$_2$  target was ablated with KrF laser radiation with a 248~nm wavelength under ultra-high vacuum \cite{19}. The film had a mirror-like surface with the rms roughness less than 12 nm, as measured by atomic force microscopy (AFM). The film thickness, $d = 90$~nm, was monitored \textit{in  situ} by a quartz balance, and finally measured by AFM and ellipsometry. The phase purity of the film was checked by X-ray diffraction and energy-dispersive spectroscopy.

Standard four-probe dc-method was used to measure resistivity at the superconducting transition. From the resistivity onset at 22~K with a transition width of 2~K we estimated $T_c = $20~K. The IR reflectivity spectra with a frequency resolution of 1~cm$^{-1}$ were measured in the 14000-8~cm$^{-1}$ range, using a gold mirror as a reference. The liquid-nitrogen cooled InSb, HgCdTe photodetectors, and the liquid-helium cooled silicon bolometer were used as detectors in the near-IR, middle-IR, and far-IR region, respectively. For measurements in the temperature range $5-300$~K, the sample was placed into the Optistat$^{\mathrm{CF-V}}$ (Oxford Instruments) cryostat with ZnSe and polyethylene windows. Wedged windows made of TPX plastic were used for measurements in the far IR range. To improve the signal-to-noise ratio, we made up to 120 runs, each of ten measurements, with a subsequent averaging.

\begin{figure}[t!]
\includegraphics[width=80mm]{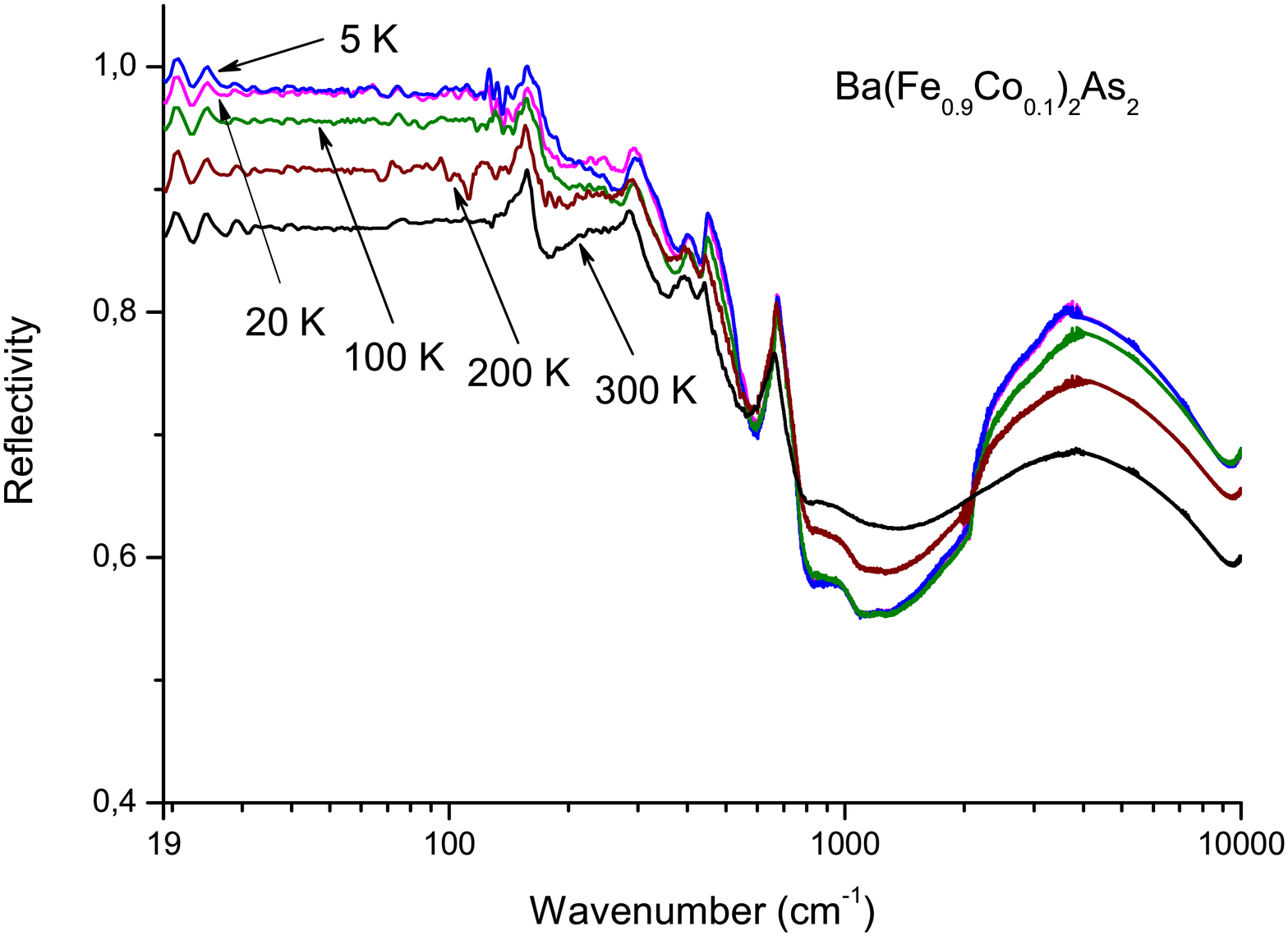}
\caption{\label{fig:pano}Infrared reflectivity spectra of the $\mathrm{Ba(Fe_{0.9}Co_{0.1})_2As_2}$ film.}
\end{figure}

%\begin{figure}[t!]
%\centering
%\epsfig{figure=fig1.eps, width=\linewidth}%\includegraphics{fig1}
%\caption{\label{fig:obz} Infrared reflectivity spectra of the %Ba(Fe$_0.9$Co$_0.1$)$_2$As$_2$ film.}
%\end{figure}

Figure~\ref{fig:pano} shows the IR reflectivity spectra of Ba(Fe$_{0.9}$Co$_{0.1}$)$_2$As$_2$ film measured at temperatures 300, 200, 100, 20, and 5~K. The small film thickness and moderate conductivity hampered calculating the conductivity and permittivity spectra by the Kramers- Kronig procedure. However, some important conclusions can be made from the straightforward analysis of the IR reflectivity spectra.

Totally, the measured IR spectra agree well with those measured earlier for the Ba(Fe$_{0.9}$Co$_{0.1}$)$_2$As$_2$ films from the same batch \cite{20}. This is also true for the far IR range where the reflectivity spectra were calculated in \cite{20} from the results of direct measurements of THz optical conductivity and permittivity. The wide band seen in Fig. 1 in the wavenumber range above 1000~cm$^{-1}$ results from the interband transitions with the maxima close to 4400 and 20800~cm$^{-1}$ \cite{20}. The dip in the region of 1000~cm$^{-1}$ can be explained by the resonance absorption. The similar feature was observed earlier in the IR reflectivity spectra for Ba(Fe$_{1-x}$Co$_x$)$_2$As$_2$ \cite{12,21,22} and also for undoped BaFe$_2$As$_2$ \cite{23,24}. It may be associated with an intraband transition \cite{25,26}. A number of the relatively narrow features in the region of 100-1000~cm$^{-1}$ is due to the phonons in the (La, Sr)(Al,Ta)O$_3$ substrate.

Our measurements reveal a new feature at $\sim$900~cm$^{-1}$ (see Fig.~\ref{fig:pano}) in the reflectivity spectra, whose origin is unknown yet. In the region below 30~cm$^{-1}$ one can see interference fringes arising due to re-reflections within the cryostat windows. This instrument effect does not hamper the analysis of the spectra.

The upturn in the reflectivity $R(\omega)$ for the wavenumbers below 1000~cm$^{-1}$ is caused by itinerant charge carriers, i.e., electrons and holes in various bands of Ba(Fe$_{0.9}$Co$_{0.1}$)$_2$As$_2$. As temperature decreases from 300 to 100~K, the reflectivity $R(\omega)$ shows a strong temperature dependence (Fig.~\ref{fig:pano}). At the same time, the spectra measured in the temperature interval $30-5$~K nearly coincide in the region of large wavenumbers, above 300~cm$^{-1}$. The difference between them appears only in the region of small wavenumbers. One can see that at temperatures lower than that of the superconducting transition for Ba(Fe$_{0.9}$Co$_{0.1}$)$_2$As$_2$, the reflectivity of the film reaches nearly unity. This provides a convincing evidence of opening of the superconducting energy gap due to formation of the SC condensate.

The opening of the superconducting gap in the electron density of states at $T<T_c$ is the basic feature of the superconducting state. In this case, for the superconducting state with the isotropic SC energy gap (the $s-$type symmetry of the order parameter), the reflectivity should reach values quite close to 100~\% at temperature 5~K $\ll T_c$ in the region of wavenumbers smaller than $2\Delta/hc$ ($c$ denotes the speed of light). The shape of the $R(\omega)$ spectra in Fig.~\ref{fig:pano} appears to be nearly flat in the region of wavenumbers smaller than 60~cm$^{-1}$ and resembles that for the superconducior with $s-$type pairing. For wavenumbers above $2\Delta/hc$ , the reflectivity decreases resulting in the formation of the characteristic peak in the frequency dependence of the ratio of reflectivities in the superconducting and normal states, $R(T<T_c)/R(T\geq T_c)$. The decrease in the amplitude of this peak in the region of small wavenumbers is due to absorption by the superconducting material at energies smaller than the superconducting gap. In the multiband superconductor, the corresponding features associated with different SC gaps coincide, thus hindering the determination of the gap magnitudes.

\begin{figure}[t!]
\includegraphics[width=80mm]{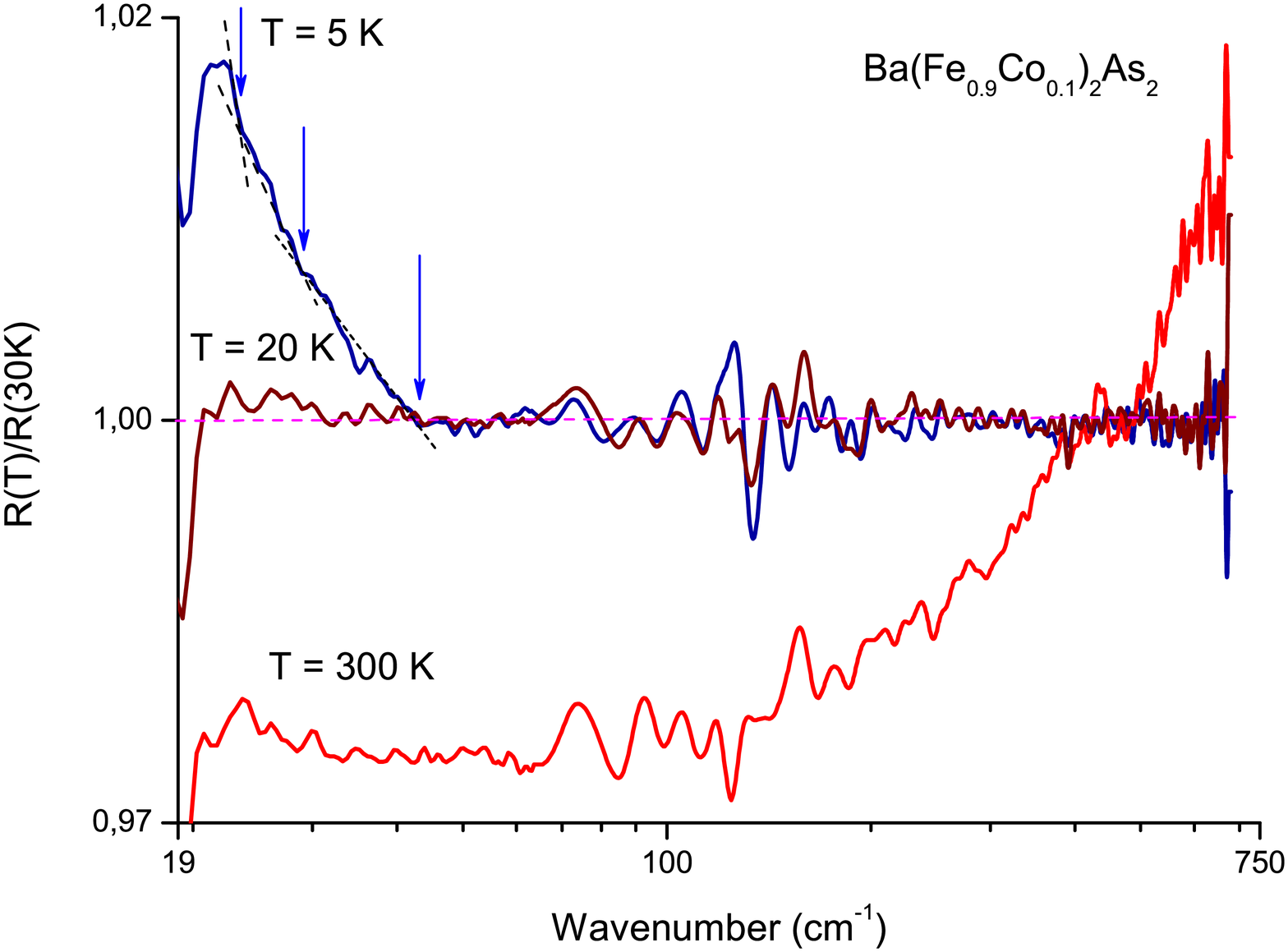}
\caption{\label{fig:norm}Infrared reflectivity spectra of the Ba(Fe$_{0.9}$Co$_{0.1}$)$_2$As$_2$ film taken at various temperatures normalized to the scectrum measured at 30 K.}
\end{figure}

Figure~\ref{fig:norm} is a semilogarithmic plot of the frequency dependence of the  $R(T)/R(30~\mathrm{K})$ ratio, where $R(T)$ is the reflectivity of the  Ba(Fe$_{0.9}$Co$_{0.1}$)$_2$As$_2$ film at temperatures $T$ = 5, 20 and 100~K. The dashed lines (drawn at the right gentle wing of the peak) show for clarity a piecewise-linear approximation of the normalized spectrum in the far IR region. One can notice the pronounced kinks in the frequency dependence of $R(5~\mathrm{K})/R(30~\mathrm{K})$ at $\sim$ 43~cm$^{-1}$ and 23.5~cm$^{-1}$ as well as a weaker feature at 29~cm$^{-1}$. The peak with the discussed features is absent in the  $R(20~\mathrm{K})/R(30~\mathrm{K})$ and  $R(5~\mathrm{K})/R(30~\mathrm{K})$ dependences, i.e. for $T > T_c$. This allows us to interpret the above features as the manifestation of the superconducting gaps, $2\Delta$ with the energies of 5.3~meV (43~cm$^{-1}$), 3.6~meV (29~cm$^{-1}$) and 2.9~meV (23.5~cm$^{-1}$). For the superconductor with $s-$type symmetry of the order parameter, the corresponding features should appear as steps in the $R_s/R_n$ dependence. However, the finite temperature of measurements and the superposition of the features related to different gaps result in smearing of the anticipated steps. It is worthy of note that in our (BG, FK, KI) previous measurements, made by terahertz spectroscopy technique \cite{20} on the samples from the same batch, the optical conductivity was found to vanish at about 30~cm$^{-1}$ and at $T = 5$~K, the fact that evidences for opening of the SC gap. On these grounds, we also interpreted the weak feature at 29~cm$^{-1}$ in the spectrum of Fig.~\ref{fig:norm} as the manifestation of the SC gap.

The fact that the SC gaps are seen in the reflectivity spectra in the far IR region suggests that the superconductivity in the studied Ba(Fe$_{0.9}$Co$_{0.1}$)$_2$As$_2$ sample corresponds to the ``dirty'' limit, i.e., the carriers scattering rate satisfies the relationship $1/\tau\geq2\Delta$. In this case, the spectral weight of the condensate in the IR reflectivity spectra and in the optical conductivity is distributed over a wide spectrum region, however, a large portion of the condensate is concentrated below the energy of the order of $2\Delta$. In the ``clean'' limit, nearly all spectral weight associated with the condensate lies below $2\Delta$, therefore at energies about $2\Delta$ no discernible changes are observed across the SC transition.

The measured SC gaps $2\Delta$ = 2.9, 3.6, and 5.3~meV correspond to the $2\Delta/k_BT_c$ ratio $\approx$ 1.6, 2.0, and 2.9, respectively. These values fall into the range of reported data $2\Delta/k_BT_c = 1.6-10$ for Ba(Fe$_{1-x}$Co$_x$)$_2$As$_2$ \cite{5,6,7,8,9,10,11,12}. To the best of our knowledge, the results reported here are the first observation of multiple superconducting gaps in the Ba(Fe$_{0.9}$Co$_{0.1}$)$_2$As$_2$ compound.

To summarize, we performed IR spectroscopic measurements on a thin film of the superconducting Ba(Fe$_{0.9}$Co$_{0.1}$)$_2$As$_2$ iron pnictide with $T_c =$~20 K in the wide spectral and temperature ranges. The behavior of the IR reflectivity spectra in the far IR range at $T<T_c$ is indicative of the $s-$type pairing in the studied material. The IR reflectivity spectra at temperatures below $T_c$ reveal the existence of three superconducting gaps $2\Delta$ = 2.9, 3.6 and 5.3~meV ($2\Delta/k_BT_c \approx$ 1.6, 2.0 and 2.9). The manifestation of the SC gaps in the IR spectra signifies the ``dirty'' limit in the studied material.

%%%\begin{acknowledgments}
This work was supported by the Russian Foundation for Basic Research, the Presidium of the Russian Academy of Sciences, the Russian Ministry of Education and Science, and the German Research Foundation (Project number HA 5934/3-1).
%%%\end{acknowledgments}


%\bibliography{apssamp}

\begin{thebibliography}{100}
\bibitem{1}Y. Kamihara, T. Watanabe, M. Hirano, and H. Hosono,
J. Am. Chem. Soc. \textbf{130}, 3296 (2008).
\bibitem{2}D. J . Singh, Physica C \textbf{469}, 418 (2009).
\bibitem{3}K. Kuroki, H. Usui, S. Onari, et al., Phys. Rev. B \textbf{79}, 
224511 (2009).
\bibitem{4}D. V. Evtushinsky, D. S. Inosov, V. B. Zabolotnyy, et al.,
Phys. Rev. B \textbf{79}, 054517 (2009).
\bibitem{5}K. Terashima, Y. Sekiba, J. H. Bower, et al., Proc. Nat.
Acad. Sci. USA \textbf{106}, 7330 (2009).
\bibitem{6}H. Ding, P. Richard, K. Nakayama, et al., Europhys.
Lett. \textbf{83}, 47001 (2008).
\bibitem{7}F. Hardy, T. Wolf. R. Fisher, et al., Phys. Rev. B \textbf{81},
060501(R) (2010).
\bibitem{8}T. J. Williams, A. A. Aczel, E. Baggio-Saitovich, et al.,
Phys. Rev. B \textbf{80}, 094501 (2009).
\bibitem{9}P. Szab\'o, Z. Pribulov\'a, G. Prist\'as, et al., Phys. Rev. B
\textbf{79}, 012503 (2009).
\bibitem{10}M. Yashima, H. Nishimura, H. Mukuda, et al., J. Phys.
Soc. Jpn. \textbf{78}, 103702 (2009).
\bibitem{11}K. Matano, Z. Li, G. L. Sun, et al., Europhys. Lett. \textbf{87},
27012 (2009).
\bibitem{12}K. W. Kim, M. R\"ossle, A. Dubroka, et al., Phys. Rev. B
\textbf{81}, 214508 (2010).
\bibitem{13}I. I. Mazin, D. J. Singh, M. D. Johannes, and M. H.
Du, Phys. Rev. Lett. \textbf{101}, 057003 (2008).
\bibitem{14}P. Samueli, Z. Pribylova, P. Szabo, et al., Physica C
\textbf{469}, 507 (209).
\bibitem{15}Yi. Yin, M. Zech, T. L. Williams, et al., Phys. Rev. Lett.
\textbf{102}, 097002 (2009).
\bibitem{16}R. T. Gordon, N. Ni, C. Martin, et al., Phys. Rev. Lett.
\textbf{102}, 127004(2009).
\bibitem{17}Y. Machida, K. Tomokuni, T. Isono, et al., J. Phys. Soc.
Jpn. \textbf{78}, 073705 (2009).
\bibitem{18}Y. Machida, K. Tomokuni, T. Isono, et al., Nature
(London) \textbf{459}, 64(2009).
\bibitem{19}K. Iida, J. H\"anish, R. H\"uhne, et al., Appl. Phys. Lett.
\textbf{95}, 192501 (2009).
\bibitem{20}B. Gorshunov, D. Wu, A. A. Voronkov, et al., Phys. Rev.
B \textbf{81}, 060509(R) (2010).
\bibitem{21}A. Dusza, A. Lucarelli, F. Pfuner, et al., Europhys. Lett. \textbf{90}, 37005 (2010).
\bibitem{22}E. van Heumen, Y. Huang, S. de Jong, et al., Europhys.
Lett. \textbf{93}, 37002 (2011).
\bibitem{23}M. Nakajima, S. Ishida, K. Kihou, et al., Phys. Rev. B
\textbf{81}, 104528 (2010).
\bibitem{24}W. Z. Hu, J. Dong, G. Li, et al., Phys. Rev. Lett. \textbf{101},
257005 (2008).
\bibitem{25}A. Kutepov, K. Haule, S. Y. Savrasov, and G. Kotliar,
Phys. Rev. B \textbf{82}, 045105 (2010).
\bibitem{26}Z. P. Yin, K. Haule, and G. Kotliar, Nature Phys. \textbf{7}, 1
(2010).
\end{thebibliography}

\end{document}